# Specific Heat Capacity of $TiO_2$ Nanoparticles

## M. Saeedian[1], M. Mahjour-Shafiei[1], E. Shojaee[1], M. R. Mohammadizadeh[1,2]

[1]*Superconductivity Research Laboratory (SRL), Department of Physics, University of Tehran, North Karegar Ave., P.O. Box 14395-547, Tehran, IRAN*
[2]*Departement of Physics, Brock University, St. Catharines, ON, L2S 3A1, Canada*

**Corresponding Author:**

**M. R. Mohammadizadeh,**

**Director, Superconductivity Research Laboratory (SRL),**
**Associate Professor of Physics,**
**Department of Physics,**
**University of Tehran,**
**North Karegar Ave.,**
**P.O. Box 14395-547**
**Tehran,**
**IRAN**
**Tel: +98 21 61118611 & 61118634**
**Fax: +98 21 88004781**

zadeh@ut.ac.ir

# Specific Heat Capacity of $TiO_2$ Nanoparticles

M. Saeedian[1], M. Mahjour-Shafiei[1], E. Shojaee[1], M. R. Mohammadizadeh[1,2]

[1]*Superconductivity Research Laboratory (SRL), Department of Physics, University of Tehran, North Karegar Ave., P.O. Box 14395-547, Tehran, IRAN*
[2]*Departement of Physics, Brock University, St. Catharines, ON, L2S 3A1, Canada*

**Abstract**

We have calculated heat capacity of $TiO_2$ nanoparticles in three stable polymorphs by applying size and surface effects on heat capacity of the bulk structure. The size and surface corrections were imposed on the acoustic and optical bulk phonons, separately. The model used in the present work is a simple modification of the model proposed by Wang *et al*. We applied the modified model to obtain the specific heat capacity of 10-100 nm $TiO_2$ nanoparticles. A very good consistency is observed between the computational and experimental data. Based on the modified model, particles with sizes larger than 70 nm behave like bulk structure. In addition, the heat capacity of particles smaller than 15 nm become independent from their structure details while demonstrating a drastic increase.

**Introduction**

Nanofluids present vast applications in modern technology. For instance, they are used for improvement of the heat transfer properties of fluids [1]. Although the nanofluid science has recently attracted great attention, but the heat transfer and specific heat capacity of nanostructures have been under study since the last decades of the twentieth century [2]. Several attempts have been made theoretically and experimentally to study the heat transfer and specific heat capacity of nanostructure materials. It has been shown that the addition of nanoparticles to the base liquid tends to enhance the heat transfer properties [1]. Titanium dioxide nanoparticles are among candidates to be used as heat capacity enhancers. The investigations of the thermal properties of $TiO_2$ nanoparticles in different sizes have been reported in the literature. Murshed *et al.* have reported the great enhancement of thermal conductivity in $TiO_2$-water based nanofluids [3]. They have measured the heat transfer of 10-40 nm titanium dioxide particles in water. Boerio *et al.* have also measured the heat capacity of 7 nm rutile and anatase nanoparticles [4]. Xin-Ming *et al.* have reported the measured heat capacity of 15 and 75 nm rutile and 14 nm anatase $TiO_2$ nanoparticles [5]. The investigations reveal the dependence of the heat capacity on nanoparticle size. More specifically, the particles with smaller size have larger heat capacity. This feature has been attributed to the surface energy, distortion and lattice energy of the nanoparticles [5].

Theoretical investigations on the thermal properties of nanoparticles are carried out in two regimes; a) small sizes (less than 10 nm), and b) large sizes where the quantum size effect corrections are negligible [6]. In the small size regime, the elastic continuum model is applicable. Baltes and Hilf [2] have derived enhancement of the specific heat by solving wave equation with free boundary conditions. They have also explained the non-quadratic behavior of the specific heat enhancement below 5 K for 22 Å lead particles, as measured by Novotny *et al.* [7]. Comparison between experimental data for Pd particles smaller than 100

Å and theoretical data, constructed based on the model proposed by Baltes and Hilf [2], has been reported to be satisfactory [8]. Another approach in the small size regiment is valence force field (VFF) model [9-11]. In the case of large size particles, where the VFF and continuum models are not applicable, some models have been proposed which impose the size and surface effects on the bulk structure properties [6]. Study of small size $TiO_2$ nanoparticles using molecular dynamics simulation has also been reported by Pavan *et al.* [11]. In the present work, we follow the method of Wang *et al.* [6] who have developed the Zhang-Benfield approach [12] to investigate the thermal properties of 10-100 nm particles. First, a review of the model and its improvements are presented and discussed. Second, the specific heat capacities of the nanoparticles are obtained and compared to the bulk and experimental data, which are available.

**Model and computational details**

Calculation of specific heat in the Zhang-Benfield approach [12] requires to know the phonon density of states (DOS) of the bulk structure. The bulk structure phonon calculations for three natural polymorphs of $TiO_2$ were carried out [13,14] in the formalism of density functional perturbative theory (DFPT) [15,16]. The calculations were performed using Quantum-Espresso package [17]. The relaxed structures of the polymorphs were used to obtain the phonon dispersion curves. The energy cutoff was 44 Ry. The k-point mesh in energy calculations was $4\times4\times4$ for all of the polymorphs, while the mesh used for calculation of the interatomic force constants was $4\times4\times4$ ($2\times2\times2$) in the case of rutile and anatase (brookite) structures [13,14]. Using phonon DOS, one can obtain the heat capacity of the bulk by the following formula:

$$C_V = \frac{1}{\rho V k_B T^2} \sum_{\vec{q},p} \frac{(\hbar\omega_{\vec{q},p})^2 e^{\hbar\omega_{\vec{q},p}/K_B T}}{\left(e^{\hbar\omega_{\vec{q},p}/K_B T} - 1\right)^2}. \tag{1}$$

Where $\rho$ is mass density of the bulk, $V$ is volume of the crystal, $q$ denotes the 1$^{st}$ Brillouin Zone (BZ) points, and $p$ enumerates the bands.

As mentioned previously, obtaining the heat capacity of the nanoparticles using the bulk phonon DOS requires imposing surface and size effects. Since the effects of size and surface are different for the acoustic and optical modes, according to Zhang and Benfiled approach [12], we have separated the heat capacity contributions of the acoustic and optical modes of the bulk structure. The heat capacity obtained solely from the acoustic mode contribution was used to obtain the velocity of sound. In order to achieve this, the Debye approximation equations,

$$C_v = 9nk_B \left(\frac{T}{\theta_D}\right) \int_0^{\frac{T}{\theta_D}} \frac{x^3 e^x dx}{(e^x - 1)^2} \tag{2}$$

and

$$v = \left(k_B \theta_D / \hbar\right) / \left(6\pi^2/\Omega\right)^{1/3} \tag{3}$$

were used. Where $\theta_D$ is Debye temperature, $n$ is number of ions in unit cell, $\Omega$ is the unit cell volume, and $v$ is sound velocity. The same goes for the Einstein model with Einstein frequency $\omega_E$ for optical modes;

$$\omega_E = k_B \theta_E / \hbar. \tag{4}$$

The specific heat capacity can be written in terms of optical ($C_{V,O}$) and acoustic ($C_{V,A}$) contributions;

$$C_V = C_{V,O}(\omega_E) + C_{V,A}(\omega = vq) \tag{5}$$

These contributions depend on $\omega_E$ and $v$ variables through which the size and surface effects are imposed. According to the Zhang and Benfiled approach the $\omega_E$ and $v$ variables are modified to implement the size and surface effects using the following variables:

$$x = \frac{N^S}{N} = \frac{36\pi}{d(b_1 + b_2 + b_3)} \tag{6}$$

$$L = \sqrt{Z^S/Z}. \tag{7}$$

Where $b_i$ are reciprocal basis vectors in the 1$^{st}$ BZ. The $N^S$ and $Z^S$ are number of surface atoms and the averaged number of bounding surface atoms, respectively. The $N$ is total number of atoms and $Z$ is the number of bounding interior atoms in the nanoparticle ($Z=N-Z^S$). The $L$ factor (Eq. 7) is a softening factor suggested by Wang *et al.* [6]. The size and surface effects on the optical phonons are imposed as following;

$$\overline{\omega}_E = xL\omega_E + (1-x)\omega_E \tag{8}$$

Therefore, one can rewrite the optical part of the specific heat of nanoparticles as;

$$C_{V,O(NP)} = \frac{(3n-3)N}{\rho V k_B T^2} \frac{(\hbar\overline{\omega}_E)^2 e^{\hbar\overline{\omega}_E/K_B T}}{\left(e^{\hbar\overline{\omega}_E/K_B T} - 1\right)^2}. \tag{9}$$

Size and surface effects on the acoustic phonons are as follow;

$$v^S = Lv \tag{10}$$

So, we can rewrite the acoustic part of the specific heat of nanoparticle, as;

$$C_{V,A(NP)} = \frac{3}{\rho V k_B T^2} \left( \sum_{\vec{q} \in Q^I} \frac{(\hbar v q)^2 e^{\hbar v q / K_B T}}{\left(e^{\hbar v q / K_B T} - 1\right)^2} + \sum_{\vec{q} \in Q^S} \frac{(\hbar v^s q)^2 e^{\hbar v^s q / K_B T}}{\left(e^{\hbar v^s q / K_B T} - 1\right)^2} \right) \quad (11)$$

where,

$$Q^I = \{\vec{q} | 0 \leq l_i \leq N_i^I\}, N_i^I = x N_i, i = 1,2,3$$

and

$$Q^S = Q - Q^I.$$

In addition, $q$ denotes the 1$^{st}$ BZ points, the $N_i$ is number of unit cells in the $i$ direction and, $N_i^I$ is number of the interior unit cells in the $i$ direction. Therefore, $Q^I$ is the set of interior $q$-points and $Q^S$ is the set of $q$-points on the surface of the 1$^{st}$ BZ. Finally, the nanoparticle heat capacity is obtained by

$$C_{V,NP} = C_{V,O(NP)} + C_{V,A(NP)}. \quad (12)$$

The above procedure was first suggested by Wang *et al.* [6]. In the present research, two main modifications were implemented in the Wang's model. The first modification is the introduction of a different softening factor. The softening factor introduced by Wang *et. al.* Eq. 7, tends to 1 as the nanoparticle size decreases. The $\varpi_E$ and $v^s$, i.e. modified Einstein frequency and modified sound velocity, tend to those of bulk values when $L$ is at its upper limit ($L \rightarrow 1$), which is an obvious failure for Eqs. 8 and 10. In order to correct this shortcoming, we have introduced the following softening factor:

$$L = 1 - \sqrt{(Z^s / Z)} \quad (13)$$

which overcomes the mentioned problem.

As the second modification, the procedure used to obtain the Debye temperature and Einstein frequency from the bulk structure phonon DOS was improved. Wang *et al*. have extracted both of them from total heat capacity of bulk material. Instead, we have deduced the Debye temperature from the acoustic phonon DOS of the bulk and the Einstein frequency from the optical phonon DOS of the bulk, separately. More specifically, we have assumed that all optical modes have an equal frequency (Einstein frequency), and acoustic branches are linear with an equal slop. The advantage of this treatment is that when we apply the size effects on the Einstein frequency (the sound velocity) we are sure that we are only modifying the optical (acoustic) modes, unlike the Wang's model. This modification has a ponderable effect on Einstein frequency (particularly at low temperatures) that is calculated merely from optical modes. In fact, the Einstein frequency obtained by our method is smaller than that obtain through Wang *et al*. procedure, as it should be.

**Results and discussion**

In this discussion we assume that the size of nanoparticles is large enough, so that the definition of 1$^{st}$ BZ is still valid. It is expected that a decrease in the size of particles (size effect) leads to a reduction of the specific heat capacity, because the number of k-points in the 1$^{st}$ BZ decreases. In the other words, there are fewer terms in the summation of heat capacity. On the other hand, transition from bulk to nanoparticle is equivalent to imposing a negative pressure on the structure which tends to soften the phonon frequency. As a consequence, an enhancement in the heat capacity should be observed (surface effect). As explained, the size and surface effects have opposite impacts on the heat capacity. Our calculations show that the surface effect is the dominant effect for TiO$_2$ nanoparticles. This result is to some extend in contradiction with the previously published article on CuO nanoparticles at low temperatures [6].

Usually, a direct comparison between experimental and theoretical volume specific heat capacity, $C_V$, is not possible. Since it is hard to measure $C_V$ experimentally, though it is rather straightforward to calculate it from the phonon spectra. In contrary to the $C_V$, it is easy to measure $C_P$, but there is no direct way to obtain it from phonon spectra. However, we know from Thermodynamics that there is a relation between $C_P$ and $C_V$ [18], so that they approach each other at low temperatures. Therefore, it is fair to replace $C_V$ by $C_P$ at low temperatures. It is what we have done while presenting our data. Since at high temperatures the effect of frequencies is suppressed, Eq. 1, the $C_V$ of nanoparticles is expected to tend to that of the bulk value (Dulang-Petitt limit).

In Fig.1, the calculated $C_V$ and the experimental $C_P$ for the bulk structure and 14 nm particles for rutile $TiO_2$ structure are shown. The calculated $C_V$ and the experimental $C_P$ for the bulk structure, and 75 (15) nm particles for rutile (anatase) structure are presented in Fig. 2 (3), as well. A common feature among these three figures is that the $C_P$ of $TiO_2$ nanoparticles crosses the Dulond-Petit limit at lower tempretuers compared to that of the bulk (It is well established that $C_P$ of the bulk must cross the Dulong-Petit limit at higher temperatures [20]). Consistently, the $C_P$ of nanoparticles crosses the Dulong-Petit limit at lower tempretures for particles with smaller size. We believe this is due to the surface effect which becomes more important for smaller particles.

Our calculations for small-size (large-size) nanoparticles are in good agreement with the experimental data below 150 K (300 K), Figs. 1 and 3 (Fig. 2). A closer look at the figures shows that the theoretical values for smaller particles (14 & 15 nm) deviate from the experimental data at a faster pace compared to that for the larger particle (75 nm) as the temperature rises. This can be attributed to the fact that experimental data are $C_P$ and theoretical values are $C_V$, whose difference is proportional to temperature and unit cell volume [20]. For smaller particles the surface effect is more important. In addition, imposing the surface effect on the polymorphs results in decreasing the phonon frequency values. As the $TiO_2$ polymorphs exhibit positive Gruneisen parameter [13,14], lowering the phonon frequency is equivalent to an increase in the unit cell volume. Consequently, the difference between $C_V$ and $C_P$ grows more rapidly as a function of temperature for smaller particles. The calculated bulk heat capacity demonstrates an excellent consistency with the experimental measurements at all temperatures.

The calculated specific heat capacities for $TiO_2$ nanoparticles of 10-100 nm in different structural phases are presented in Figs. 4 to 6. As seen, the specific heat of nanoparticles always stands above that of the bulk at all temperatures. Unfortunalty, there is no report, to date, to compare our calculations with $C_V$ of brookite $TiO_2$ nanoparticles. The calculated volume specific heat capacities for particles for the three polymorphs of 10 to 100 nm at room temperature are presented in Fig. 7. It seems that the particles larger than about 70 nm behave like bulk structure. It can be seen that the specific heat capacity values corresponding to different polymorphs approach each others as demonstrating a fast increase while the particle size drops below 15 nm. This observation suggests that the microstructure in the small sizes is no longer relevant. We believe, at small sizes the surface atoms contribution to the heat capacity becomes very large, and since these atoms do not play a serious role in the polymorph structure, the heat capacity at small sizes becomes independent from the polymorph structure. Therefore, it is recommended to use $TiO_2$ particles with sizes smaller than 15 nm, regardless of the polymorphs, in order to improve nanofluidic heat capacity property.

**Conclusions**

The model proposed by Wang *et al.* was modified to investigate the specific heat capacity of three stable polymorphs of $TiO_2$ nanoparticles in the range of 10-100 nm size. Comparison with the available experimental data shows good agreements in the case of rutile and anatase nanoparticles. This model is a free-parameter model, and only needs the bulk phonon density of states to obtain the specific heat of nanoparticles. Based on this modified model, the particles with sizes larger than 70 nm behave like bulk structure. In addition, a drastic increase in the specific heat capacity of particles smaller than 15 nm, independent from their microstructure, is observed.

**Acknowledgments**

The partial financial support by the research council of the University of Tehran is acknowledged. Technical support of the computational nanotechnology supercomputing center at the Institute for research in the fundamental sciences (IPM), S. Mohammadi, M. F. Miri, and M. Abbasnejad are acknowledged. One of the authors (M. R. Mohammadizadeh) would like to appreciate F. Razavi's support during his sabbatical in Brock University.

# References

**1.** Y. Xuan and Q. Li, Int. J. Heat Fluid Flow. 21, 58 **(2000)**.

**2.** H.P. Baltes and E.R. Hilf, Solid State Commun. 12, 369 **(1972)**.

**3.** S.M.S. Murshed, K.C. Leong, and C. Yang, Int. J. Therm. Sci. 44, 367 **(2005)**.

**4.** J. Boerio-Goates, G. Li, L. Li, Trent F. Walker, T. Parry, and B. F. Woodfield, Nano Lett. 6, 750 **(2006)**.

**5.** W. Xin-Ming, W. Lan, T. Zhi Cheng, H. L. Guang, and S.Q. Song, J. Solid State Chem. 156, 220 **(2001)**.

**6.** B. Wang, L. Zhou, and X. Peng, Int. J. Thermophys. 27, 139 **(2006)**.

**7.** V. Novotny, P.P. M. Meincke, and J.H.P. Watson, Phys. Rev. Lett. 28, 901 **(1971)**.

**8.** G.H. Comsa, D. Heitkamp, and H.S. Rade, Solid State Commun. 24, 547 **(1977)**.

**9.** W.A. Harrison, *Electronic Structure and the Properties of Solid, Freeman*, San Francisico, **(1980)**, Chapter 4

**10.** S.F. Ren and W. Cheng, J. Appl. Phys. 100, 103505 **(2006)**.

**11.** N.K. Pavone, P.T. Cummings, H. Zhang, and J.F. Banfield, J. Phys. Chem. B 109, 15243 **(2005)**.

**12.** H. Zhang and J. F. Benfield, Nano Structured Materials 10, 185 **(1998)**.

**13.** E. Shojaee and M. R. Mohammadizadeh, J. Phys.: Condens. Matter 22, 015401 **(2010)**.

**14.** E. Shojaee, M. Abbasnejad, M. Saeedian, and M. R. Mohammadizadeh, Phys. Rev. B Accepted for publication **(2011)**.

**15.** P. Giannozzi, S. de Gironcoli, P. Pavone, and S. Baroni, Phys. Rev. B 43, 7231 **(1991)**.

**16.** X. Gonze and C. Lee, Phys. Rev. B 55, 10355 **(1997)**.

**17.** P. Giannozzi, S. Baroni, N. Bonini, M. Calandra, R. Car, C. Cavazzoni, D. Ceresoli, G. L. Chiarotti, M. Cococcioni, I. Dabo, A. Dal Corso, S. de Gironcoli, S. Fabris, G. Fratesi, R. Gebauer, U. Gerstmann, C. Gougoussis, A. Kokalj, M. Lazzeri, L. Martin-Samos, N. Marzari, F. Mauri, R. Mazzarello, S. Paolini, A. Pasquarello, L. Paulatto, C. Sbraccia, S. Scandolo, G. Sclauzero, A.P. Seitsonen, A. Smogunov, P. Umari, and R.M. Wentzcovitch, QUANTUM ESPRESSO: a modular and open-source software project for quantum simulations of materials, J. Phys.: Condens. Matter 21, 395502 **(2009)**.

**18.** D. de Ligny, E.P. Richet, E.F. Westrum Jr. and J. Roux, Phys. Chem. Minerals, 29, 267 **(2002)**.

**19.** S. J. Smith and R. Stevens, American Mineralogist 94, 236 **(2009)**.

**20.** M.W. Zemansky and R.H. Dittman, *Heat and Thermodynamics, Six Edition*, McGraw-Hill, **(1981)**, Chapter 9

**Figures Captions**

**Fig. 1** Calculated specific heat capacity of 14 nm particles and the bulk structure for rutile $TiO_2$ compared to the experimental data [5,18].

**Fig. 2** Calculated specific heat capacity of 75 nm particles and the bulk structure for rutile $TiO_2$ compared to the experimental data [5,18].

**Fig. 3** Calculated specific heat capacity of 15 nm particles and the bulk structure for anatase $TiO_2$ compared to the experimental data [5,19].

**Fig. 4** Calculated specific heat capacity for 10 up to 100 nm rutile particles in terms of temperatures compared to the bulk values.

**Fig. 5** Calculated specific heat capacity for 10 up to 100 nm anatase particles in terms of temperatures compared to the bulk values.

**Fig. 6** Calculated specific heat capacity for 10 up to 100 nm brookite particles in terms of temperatures compared to the bulk values.

**Fig. 7** Calculated specific heat capacity for 10 up to 100 nm $TiO_2$ nanoparticles at 300 K, as function of particle size.

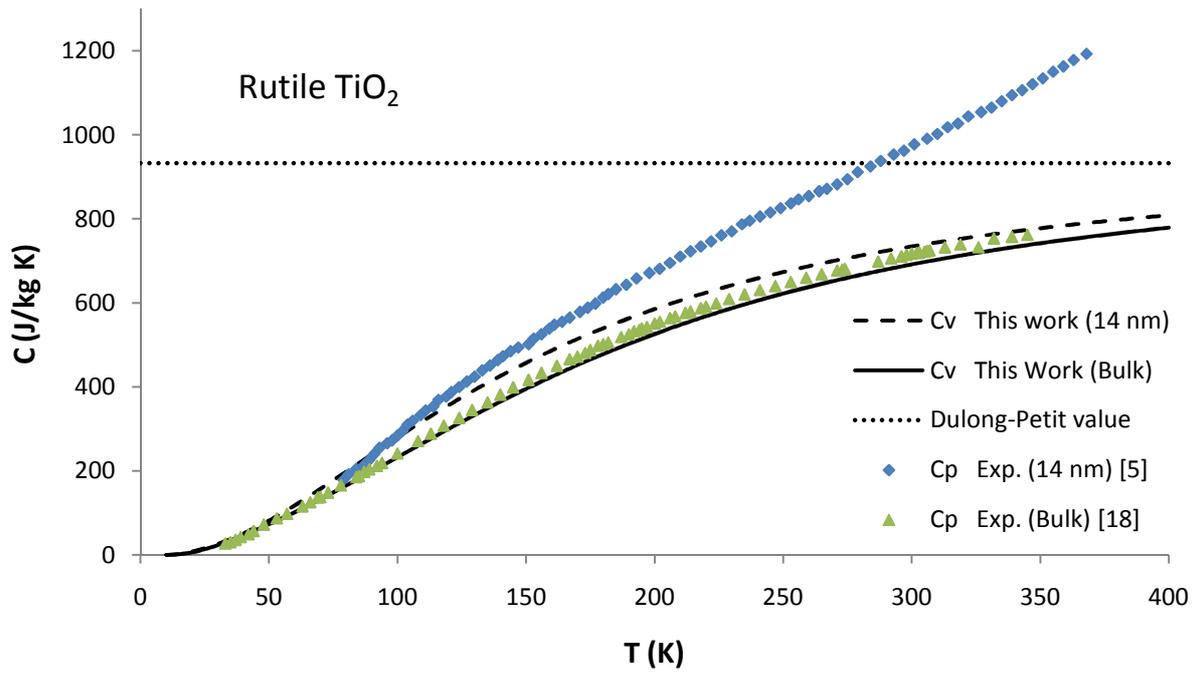

**Fig. 1**

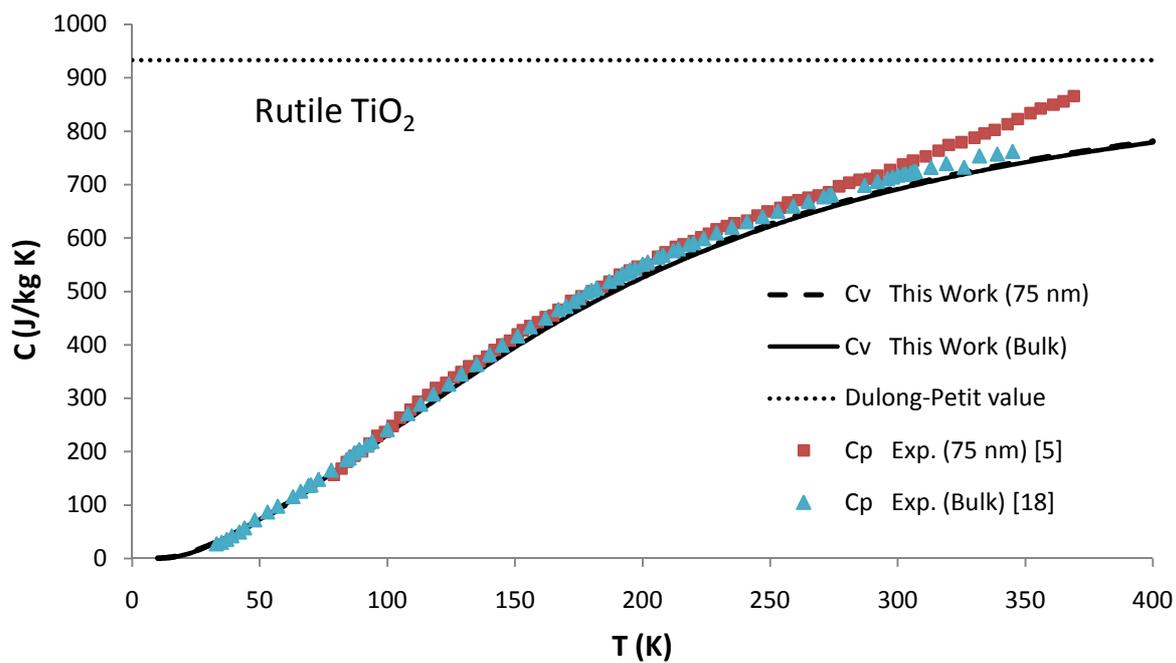

**Fig. 2**

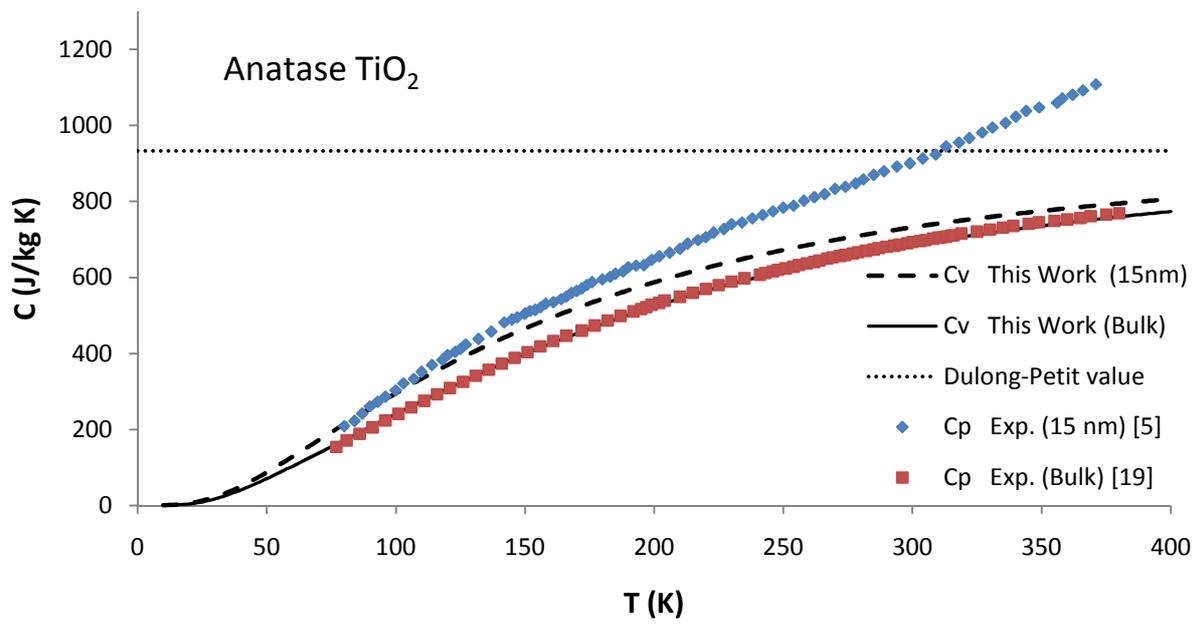

**Fig. 3**

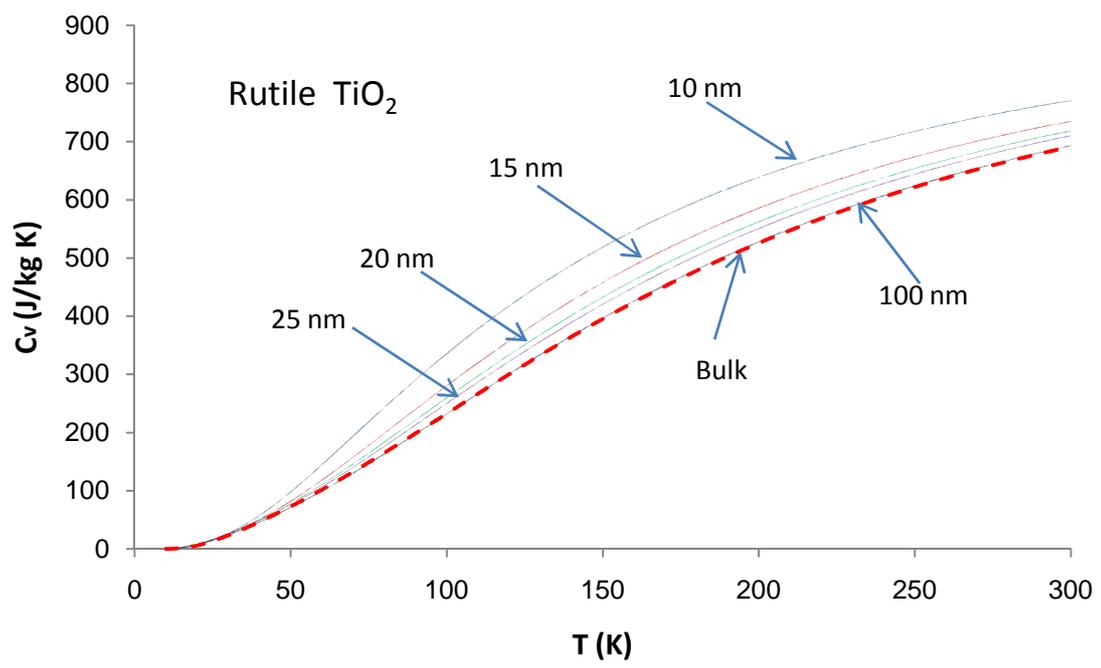

**Fig. 4**

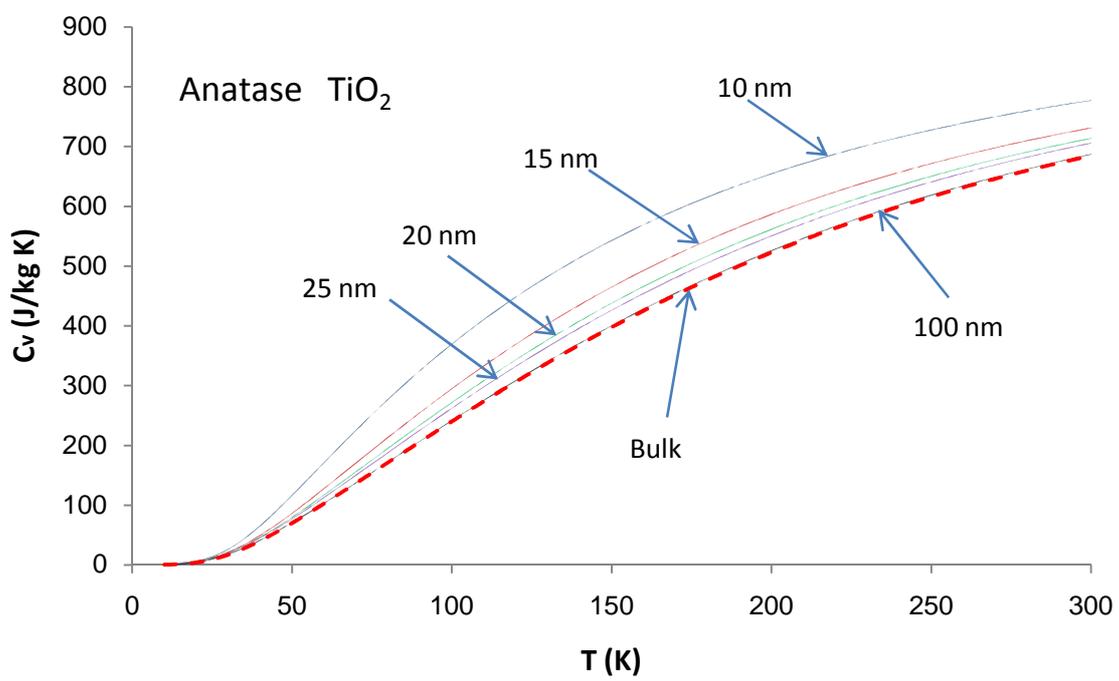

**Fig. 5**

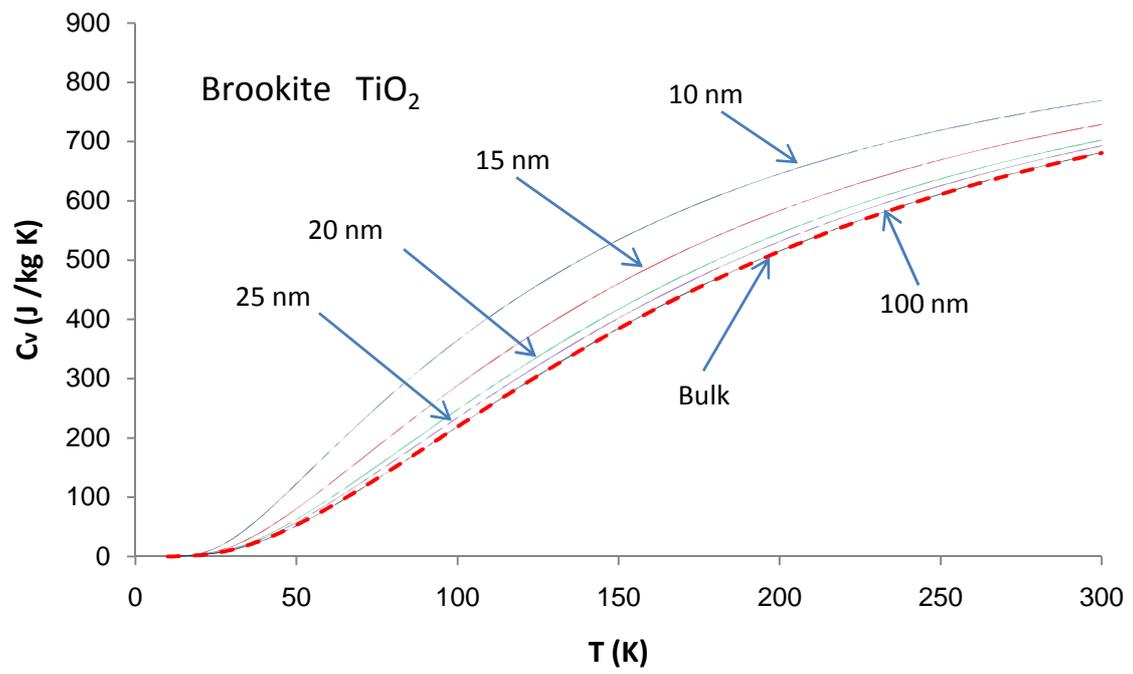

**Fig. 6**

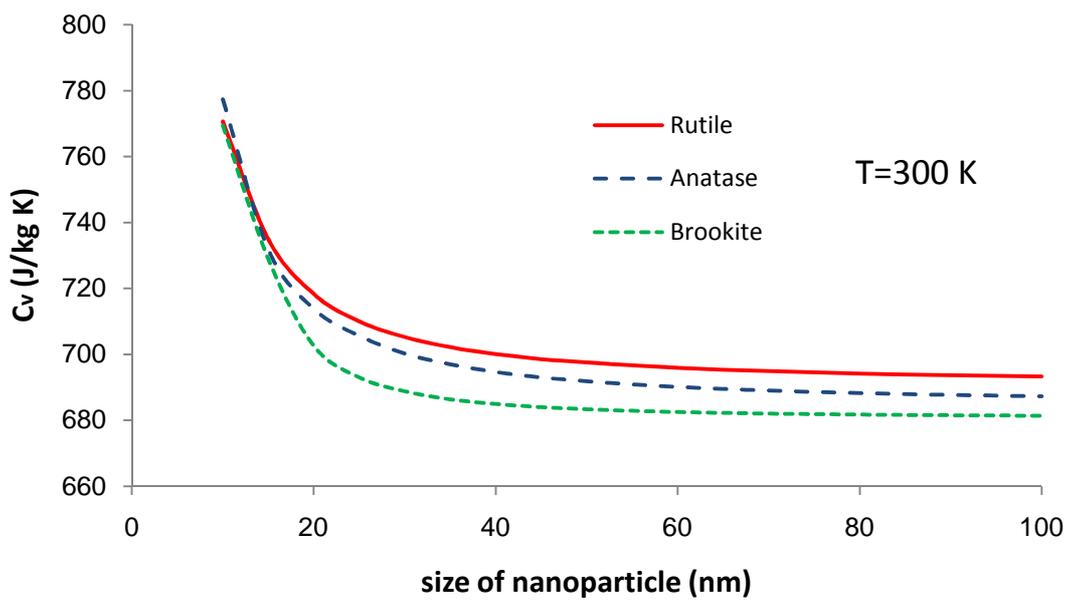

**Fig. 7**

| | |
|---|---|
| Filename: | Nanopaper.docx |
| Directory: | C:\Users\Meghdad\Documents |
| Template: | C:\Users\Meghdad\AppData\Roaming\Microsoft\Templates\Normal.dotm |
| Title: | |
| Subject: | |
| Author: | PARAND |
| Keywords: | |
| Comments: | |
| Creation Date: | 4/14/2010 11:00:00 AM |
| Change Number: | 367 |
| Last Saved On: | 4/23/2011 10:09:00 AM |
| Last Saved By: | PARAND |
| Total Editing Time: | 3,542 Minutes |
| Last Printed On: | 7/29/2013 4:38:00 PM |
| As of Last Complete Printing | |
|    Number of Pages: | 21 |
|    Number of Words: | 3,146 (approx.) |
|    Number of Characters: | 17,933 (approx.) |